\documentclass[letter]{article}
\usepackage{fullpage}
\usepackage{url}
\usepackage{graphicx}
\usepackage{mathtools}
\usepackage{subcaption}
\usepackage[usenames, dvipsnames]{color}
\DeclareGraphicsExtensions{.eps,.pdf,.jpeg,.png}

\usepackage[hidelinks]{hyperref}
\usepackage{cite}
\usepackage{breakurl}

\title{System-level Scalable Checkpoint-Restart \\ for Petascale Computing}

\author{
Jiajun Cao$^1$\thanks{$^*$ This work was partially supported
            by the National Science Foundation under Grant~ACI-1440788.}
\and Kapil Arya$^2$ 
\and Rohan Garg$^1$$^*$ 
\and Shawn Matott$^3$ 
\and\hskip-10pt Dhabaleswar K. Panda$^4$
\and\hskip-10pt Hari Subramoni$^4$
\and\hskip-10pt J\'er\^ome Vienne$^5$
\and\hskip-10pt Gene Cooperman$^1$$^*$\\
\small
\and $^1$Northeastern University\\
     Boston, MA\\
    \{jiajun,rohgarg,gene\}@ccs.neu.edu
\and $^2$Mesophere, Inc.\\
     San Francisco, CA\\
     kapil@mesosphere.io
\and $^3$CCR: Center for Computational Research\\
     State University of New York at Buffalo\\
     lsmatott@buffalo.edu
\and $^4$The Ohio State University\\
     Columbus, OH\\
     \{panda,subramon\}@cse.ohio-state.edu
\and $^5$Texas Advanced Computing Center\\
     The U. of Texas at Austin\\
     viennej@tacc.utexas.edu
}

\date{}

\begin{document}

\maketitle

\begin{abstract}
Fault tolerance for the upcoming exascale generation has long been an area
of active research.  One of the components of a fault tolerance strategy
is checkpointing.  Petascale-level checkpointing is demonstrated through
a new mechanism for virtualization of the InfiniBand~UD (unreliable
datagram) mode, and for updating the remote address on each UD-based
send, due to lack of a fixed peer.  Note that InfiniBand~UD is required
to support modern MPI implementations.
An extrapolation from the current results to future SSD-based storage
systems provides evidence that the current approach will remain practical
in the exascale generation.
This transparent checkpointing approach is evaluated using a framework of the
DMTCP checkpointing package.  Results are shown for HPCG (linear algebra),
NAMD (molecular dynamics), and the NAS NPB benchmarks.  In tests up to
32,752 MPI processes on 32,752 CPU cores, checkpointing of a computation
with a 38~TB memory footprint in 11~minutes is demonstrated.
Runtime overhead is reduced to less than 1\%.
The approach is also evaluated across three widely used MPI
implementations.
\end{abstract}

\section{Introduction}

Scalability of checkpointing for petascale and future exascale computing
is a critical question for fault tolerance on future
supercomputers.  A stream of publications by researchers has
been concerned with this question of fault tolerance for future
supercomputers~\cite{elnozahy2004checkpointing,cappello2009fault,
cappello2009toward,cappello2014toward,snir2014addressing}.

System-level transparent checkpointing has been avoided at larger scale in
HPC because of the need for a full-memory dump.  For example, a
2014~report~\cite{cappello2014toward} on software resilience presents
a typically pessimistic outlook for pure full-memory checkpoints:
\begin{quotation}
\noindent
``The norm in 2009 was to store the application state on remote storage,
generally a parallel file system, through I/O nodes. Checkpoint time was
significant (often 15–-30 minutes), because of the limited bandwidth
of the parallel file system. When checkpoint time is close to the MTBF,
the system spends all its time checkpointing and restarting, with little
forward progress. Since the [MTBF] may be an hour or less on exascale
platforms, new techniques are needed in order to reduce checkpoint
time.''~\cite{cappello2014toward}
\end{quotation}

Nevertheless, prior work on transparent, system-level checkpointing
is only used at moderate-scale (e.g., 128~MPI processes
in~\cite{liu2015elastic,shahzad2013evaluation}).
The single-node checkpointing
package BLCR~\cite{BLCR03,BLCR06} is used in combination with
the checkpoint-restart service of a given MPI implementation such
as~\cite{hursey2009interconnect} for Open~MPI or \cite{sankaran2005lam}
for LAM/MPI. In this approach, the checkpoint-restart service temporarily
tears down the InfiniBand network, and delegates the single-process
checkpointing to BLCR. This approach does not scale, since BLCR does
not support SysV shared memory objects~\cite{blcr-limitations}. Most
modern MPI implementations require such shared memory for efficient
communication among MPI processes on the same node — to avoid the
delay in going through kernel system calls.

Moreover, an important work on transparent, system-level checkpointing
is~\cite{cao2014transparent}, which supported only InfiniBand~RC
(reliable connection)
mode.  While that result sufficed for earlier MPI implementations, modern
MPI implementations require InfiniBand~UD for optimal performance
when running with more than about 64~processes.  This is because a
pure point-to-point RC~mode implementation would require up to $n^2$
connections for $n$~MPI ranks (for $n$~MPI processes).  MPI requires
InfiniBand~UD for the scales considered in this work, such as~32,752 MPI
processes on 32,752 CPU cores.  Setting up $(32,752)^2$, or nearly 1~billion,
point-to-point RC connections is unacceptable both due to large memory
resources and long times for initialization.

Advances in transparent checkpointing on large-scale
supercomputers depend on the fundamental problem of transparent
checkpointing over InfiniBand:  how to save or replay ``in-flight data''
that is present in the InfiniBand network at the time of checkpointing,
while at the same time not penalizing standard runtime performance.
In particular, we need to address (a)~how to enable transparent
checkpointing support for InfiniBand UD (unreliable datagram)
mode; and (b)~how to reduce the excessive runtime overhead seen
at larger scales.  This second issue of runtime overhead
affects performance even when no checkpoints are taken.  The earlier
result~\cite{cao2014transparent} had shown runtime overhead to
grow as high as 1.7\% ith 2K~cores.  When scaling to 4K~cores
on the Stampede supercomputer in this work, overhead then grew
to an unacceptable~9\% (see Section~\ref{sec:runtime-overhead} for a
discussion and solution).

\subsection{Contributions}
\label{sec:contributions}
The primary contribution of this paper is to demonstrate the practicality
of petascale system-level checkpointing through the use
of full-memory dumps.
In order to achieve this, DMTCP~\cite{ansel2009dmtcp} was used as a
vehicle for checkpointing.
We have extended the designs for DMTCP software to have
a tight interaction with modern MPI runtimes by taking
advantage of some important scalability features. The proposed enhancements are
along these directions:

\begin{enumerate}
\item This is the first checkpoint support for a hybrid InfiniBand
  communication mode that uses both reliable connection (RC) and
  unreliable datagram (UD). A hybrid RC/UD~mode provides better
  performance than a pure connection-oriented RC~mode, and is a
  commonly used optimization among modern MPI implementations.
  See Section~\ref{sec:udModel} for details.

\item A secondary contribution is to lower the runtime overhead
	for checkpointing RC~mode connections themselves.  The previous work
	supported only RC~mode~\cite{cao2014transparent}, using
	runtime tracing of InfiniBand send messages. The runtime
        overhead was shown to be 1.7\% for 2K~cores (see Table~2
        in~\cite{cao2014transparent}), which
	grew to 9\% for 4K~cores in current experiments on Stampede.
        We use a new checkpoint-time strategy
        that reduces runtime overhead to under~1\% even for many cores
        (see Section~\ref{subsec:runtime}).

\end{enumerate}

\subsection{Significance of this Work}

The current work represents an advance in the state-of-the-art.
By transparently supporting both InfiniBand RC and UD mode,
this work demonstrates a
pure system-level checkpoint over 32,752 CPU cores at the petascale
Stampede
supercomputer~\cite{stampede} in just 10.9~minutes,
during which 38~TB are saved to stable storage on a Lustre filesystem.
In contrast, an earlier report~\cite{cappello2014toward} described the 2009
state-of-the-art for checkpointing to be 15--30 minutes for a
supercomputer from that era, and had argued that checkpointing
times would increase even further from the times of that era.

Of course checkpointing by creating full-memory dumps is just one
component of a software resilience strategy of the future, and is
compatible with other complementary approaches.  These include multi-level
checkpointing~\cite{moody2010design}, incremental checkpointing,
partial restart, mitigation of silent data corruption (SDC), tuning
of checkpoint frequencies~\cite{liu2008optimal,tiwari2014lazy},
and alternate approaches to error prevention, prediction,
tolerance, detection, containment and recovery (forward or
backward)~\cite{cappello2014toward,snir2014addressing}.

\subsection{Going beyond the petascale HPC systems}
Going beyond the petascale level presented here, there is an important
question of scalability to support future exascale computing.  In order
to address this, we enunciate a simple formula, the Checkpoint Fill-Time
Law, for predicting the checkpoint time using fundamental specifications
for a given supercomputer (see Section~\ref{sec:ssd}).  This law predicts
an ideal checkpoint time, and underestimates the checkpoint time for two
real-world applications (HPCG and NAMD) by as much as a factor of about ten.
Nevertheless, this formula predicts that SSD-based exascale supercomputers
of the future will enable checkpointing through a full-memory dump in just
1.6~minutes (ideally), or a real-world 16~minutes if one extrapolates
using the same factor of ten that is seen at the petascale level.

In order to gain confidence in the predictions for an SSD-based supercomputer,
we also tested on a single SSD-based computer in Table~\ref{tbl:ckptFillTime}.
A 3~GB checkpoint image was created there in 7.2~seconds (and restart required
6.2~seconds).  This is an I/O bandwidth of 416~MB/s, close to the ideal
bandwidth of 500~MB/s for SATA 3.0 interface.  Since 3~GB is 2.3\% of the 128~GB
SSD disk, the predicted ideal checkpoint time is 2.3\% of 4.3~minutes,
or 5.9~seconds.  So, the predicted time of 5.9~seconds compares well
with the actual time of 7.2~seconds.

\subsection{A Remark on the Use of TCP on a Supercomputer}
In addition to the research contributions above, we were surprised
to discover a counter-intuitive practical issue in checkpointing at
petascale levels.  Simply launching a new computation was found to
be excessively slow with 8K~cores, and was found to fail at 16K~cores
(see Section~\ref {sec:socketLimitations}).  This was tracked down to
limitations in the hardware/software system.  The simple act of creating
16K~TCP sockets (from each process to the coordinator) was found to
overwhelm the hardware/software system on the Stampede supercomputer.
In discussions with sysadmins, we found that in the emphasis on InfiniBand
over Ethernet meant that each rack at Stampede was provided with just a
single 10~Gb~Ethernet backbone from each rack.  Hence, this appears to
have led to longer delays in the processing of Ethernet by the kernel
at larger scales, and we directly observed processes receiving a SIGKILL
signal from the kernel at 16K~cores.

\subsection{Organization of Paper}
The rest of this paper is organized as follows.
The relevant background on Lustre, DMTCP, and the various modes
used by MVAPICH2 are
presented in Section~\ref{sec:background}.
Section~\ref{sec:implementation} describes the methodology used to achieve
petascale level and some implications for extending checkpointing
to the next level.
The experimental evaluation is presented in Section~\ref{sec:experiment}.
Section~\ref{sec:discussion} describes the scalability issues
associated with petascale checkpointing.
The related work is presented in Section~\ref{sec:related},
and conclusions appear in Section~\ref{sec:conclusion}.

\section{Background}
\label{sec:background}
The following three subsections review three critical components that
affect the performance in the experiments: the MPI implementation
(MVAPICH2 at TACC, and Intel~MPI and Open~MPI at CCR --- see
Section~\ref{sec:experiment}), DMTCP itself as the checkpointing software,
and Lustre as the back-end filesystem.

\subsection{MVAPICH2}
\label{sec:mvapich2}
We highlight MVAPICH2~\cite{mvapich2} as the MPI used in the majority
of experiments.  Other MPI implementations typically have similar
features to those described here.  MVAPICH2 uses the TCP/IP-based Process
Management Interface (PMI) to bootstrap the InfiniBand end-points using
InfiniBand~RC.  While PMI is the most straightforward way to establish
InfiniBand connectivity, it leads to poor startup performance due to
the $n^2$ point-to-point connections referred to in the introduction.
For MPI jobs with more than 64~processes, MVAPICH2 also
uses the lazy establishment
of ``on-demand'' connections using InfiniBand~UD~\cite{chakrabs:hips15}
(although the 64-process threshold can be configured using the {\tt
MV2\_ON\_DEMAND\_THRESHOLD} environment variable).

\subsection{DMTCP}
\label{sec:backgroundDMTCP}
Distributed MultiThreaded
CheckPointing (DMTCP)~\cite{ansel2009dmtcp} provides a framework for
coordinated checkpointing of distributed computations via a centralized
coordinator.  Each client process of the application communicates with
the coordinator via a TCP socket.

DMTCP includes a checkpointing library that is injected into each
process of the target application. This library creates a checkpoint
thread in each process, to communicate with the coordinator and
to copy process memory and other state to a checkpoint image.

The coordinator implements global barriers to synchronize
checkpoint/restart between multiple nodes, and it provides a
publish-subscribe scheme for peer discovery (e.g., discover new TCP
peer addresses for InfiniBand id during restart).  These are used in
combination with wrappers around library functions to build plugin
libraries.  The plugin libraries are injected along with the checkpoint
library.  They serve to translate real ids into virtual ids seen by the
application, and to update the virtual address translation table with
the new real ids that are seen on restart~\cite{arya2016design}.
This virtualization capability is used to virtualize below the level
of the MPI library~\ref{subsec:ompiAndImpi}).  A new plugin capability
for this work serves to virtualize the InfiniBand~UD mode.

\subsection{Lustre}
\label{sec:lustre}
The Lustre filesystem at Stampede plays a critical role in supporting
high-bandwidth writes of checkpoint image files.  Lustre~\cite{lustre} is
a parallel object-based filesystem in widespread use that was developed
to support large-scale operations on modern supercomputers. Lustre
attains high I/O performance by simultaneously striping a single
file across multiple Object Storage Targets (OSTs) that manage the
system's spinning disks. Lustre clients run the Lustre file system and
interact with OSTs for file data I/O and with metadata servers (MDS)
for namespace operations. The Lustre protocol features authenticated
access and write-behind caching for all metadata updates.

\section{Issues for Petascale Checkpointing and
	Extrapolation to Exascale Checkpointing}
\label{sec:implementation}

In the first subsection, we discuss a key barrier to
petascale checkpointing and its solution:  support for
InfiniBand UD~mode.  In the nature of lessons learned, we also present two
additional and unexpected barriers to scalability
within the context of running on the Stampede supercomputer:
excessive runtime overhead at large scale, and the lack
of support for processes employing many TCP sockets.

Finally, the scalability of this approach for future exascale
supercomputers is a key concern.  The key question here is the write
bandwidth to the storage subsystem for a full-memory dump from
RAM.  Section~\ref{sec:ssd} presents a simple, empirical model, the
Checkpoint-Fill-Time Law, to extrapolate trends, and predicts that with
the adoption of SSD for secondary storage in supercomputers (and with
hard disks being relegated to tertiary storage), expected checkpoint
times in the exascale generation are estimated at 1.6~minutes, ideally,
and 16~minutes in real-world applications.

\subsection{Checkpointing Support for UD (Unreliable Datagrams)}
\label{sec:udModel}

Recall from Section~\ref{sec:mvapich2} that the InfiniBand UD communication mode
is for connectionless unreliable datagrams.
Newer versions of MPI use a hybrid RC/UD scheme for
balancing performance with the memory requirements for
the queue pairs.  Thus, transparent checkpointing of modern MPI
requires support for UD and in particular for hybrid RC/UD
mode (in which both types of communication operate in parallel).

\begin{figure}[ht]
\centering
\includegraphics[width=0.7\columnwidth]{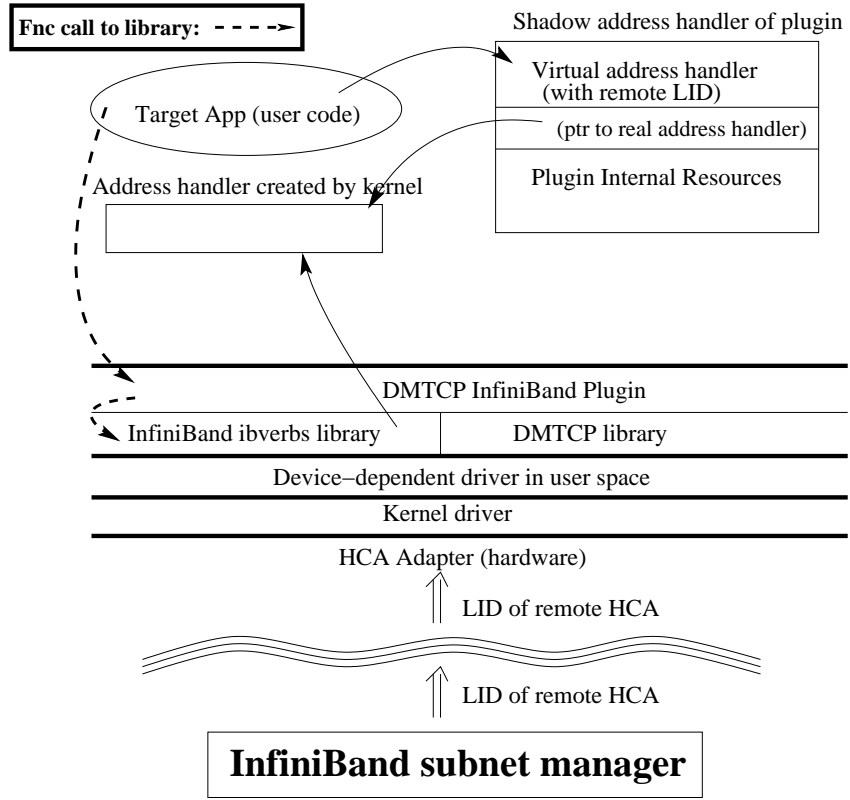}
\caption{\label{fig:ibv-ud-ah} Virtualization of 
	address handler and LID (local id) of remote HCA (hardware
	channel adapter).}
\end{figure}

The key to checkpointing UD is to virtualize remote address of the queue
pairs, so that the actual address can be replaced by a different
address after restart.  Figure~\ref{fig:ibv-ud-ah} presents an
overview of the situation, to accompany the detailed description
that follows.

The approach here maintains a translation
table between virtual and actual addresses, and is implemented
using DMTCP plugins~\cite{arya2016design}.
Further, on each UD-based send, the InfiniBand LID (local identifier)
must also be updated for a possibly different remote
queue-pair address.

In detail,
each computer node includes an InfiniBand HCA (Host Channel Adapter).
The HCA provides hardware support for a {\em queue pair}, which refers
to a send-receive pair of queues.  Unlike the connection-oriented RC
communication mode, UD has no end-to-end connections.  So a local
queue pair can send messages to different remote queue pairs.

The problem of virtualizing a remote UD address is made more difficult
because of the dynamic change of the address of the remote queue pair,
which is identified by a unique pair (LID (local identifier), qp\_num
(queue pair number)). The LID is assigned by the subnet manager, which
is unique only within the local subnet, while the queue pair number is assigned
by the hardware driver, which is unique only within the local HCA. Since
both fields can change after restart, we need to virtualize both fields
in order to identify the remote UD address.

At the time of restart, all previously created UD queue pairs (as well
as the ones created after restart) will send their address pairs to
the DMTCP checkpoint coordinator. After the application resumes, each node
must discover the (remote-LID, queue-pair-number).
It was decided to do this by querying the checkpoint coordinator at runtime
prior to each UD-based send for the latest destination LID.
Although this adds to the runtime overhead,
UD is not used as the primary communication
channel by MPI implementations,
and so this runtime querying overhead is negligible in practice.

Note that UD presents a very different situation from the older RC~mode
work in~\cite{cao2014transparent}.  For RC~mode it's safe to build
the connection at restart, because the peer won't change.  But the
destination of a given LID can change after restart.  In the extreme
case, each UD-based send can change its peer queue-pair address, and so
there's no fixed peer.  Instead, we are forced to patch the right remote
address on every send.

Finally, the UD protocol refers to an address handler (for which the
remote LID is just one field).  So, instead of virtualizing the remote LID,
we must virtualize the address handler (AH).
Hence, we create wrapper functions around
all InfiniBand library calls that refer to an AH.  Whenever an AH
is created, we also create a shadow AH.  Thus, the application code
only sees pointers to our shadow AH, and our wrapper functions
make corresponding modifications to the actual AH struct that is
passed to the InfiniBand library.  On restart, the shadow AH is
redirected to a new, actual AH constructed by the InfiniBand library.
In particular, this technique can account for any hidden fields
in the actual AH, or other examples of data hiding that an InfiniBand
library may use.

\setcounter{paragraph}{0}

\subsection{Reducing RC-mode Runtime Overhead}
\label{sec:runtime-overhead}

In testing on the LU.E benchmark,
we saw runtime overhead rise to 9\% for 4K~CPU cores compared to the 1.7\%
runtime overhead at 2K~cores reported by~\cite{cao2014transparent}.
This was due to the non-scalable tracing of send/receive requests required
by the InfiniBand checkpointing code to shadow hardware state, since
InfiniBand devices don't provide a way to ``peek'' at the current state.

We address this scaling problem by updating the model by
relaxing some of the guarantees around send/receive queues.
Instead of computing the exact number of pending send messages at
checkpoint time, we poll the receive queues for
a ``reasonable''
amount time during checkpointing. If a message arrives during this time,
we wait again. If no messages arrive, we assume that there
are no more messages in flight. For practical purposes, most message
will arrive in the first time window. There might be a small number of
messages arriving in the second time window if there is a slow network switch.
Since the InfiniBand network is quiesced at this point (because
all processes are going through checkpoint), no new messages are being
scheduled to send. In our experiments, we used a ``one-second window'' for
draining in-flight messages and noticed that all ``pending''
messages arrived within the first window. No messages arrived in the
second window.

\subsection{TCP Congestion Limits during MPI Initialization}
\label{sec:socketLimitations}

While startup time was reasonable for DMTCP with 8K~clients, when
running with 16K~clients, some of the clients randomly died.
We were able to diagnose this by creating an artificial application
unrelated to DMTCP.  The application similarly created a TCP
connection between a coordinator and each client.  We observed
a SIGKILL being sent to each of the random processes that died.
Since standard user-space tools failed to indicate the sender of the SIGKILL,
The behavior was reproducible:  DMTCP ran well with 8K~clients, but was
never observed to run with 16K~clients.

In discussions with the sysadmins, they pointed out that there was
a single 10~Gb~Ethernet backbone from each rack, since the cluster
emphasized InfiniBand over Ethernet.  we speculate that the Linux kernel
had sent the SIGKILL due to excessive delays seen by the kernel on top
of an overloaded Ethernet backbone.

We then implemented two standard solutions.  First, a ``staggered
sleep'' (i.e., network backoff) was used to avoid bursts of communication
over the network when initializing the TCP sockets.  This worked.
However, Stampede also sets a per-user limit of 16K sockets per process
(which can be individually overridden by the system administrator on a
per-user basis).

So, in order to scale to 16K~MPI processes and beyond, we then employed
a second solution.  We created a new mode using a two-level tree of
coordinators.  Specifically, a ``sub-coordinator'' process was created on
each computer node to relay messages to the main coordinator.  In certain
cases, the messages of clients were aggregated by the sub-coordinator
into a single message in order to optimize network overhead.  As shown
in section~\ref{tbl:launch}, the launch time improved significantly when
using this two-level tree of coordinators.

\subsection{SSDs as a Disk Replacement in the Exascale Generation}
\label{sec:ssd}

As is well known, the bottleneck for a transparent checkpoint employing
a full-memory dump is the sustained write bandwidth (sustained transfer
rate) to the storage subsystem.  This bears on the practicality of
this approach.  We argue here through a crude model that transparent
checkpointing through full-memory dumps will actually become faster in
the exascale generation if exascale computers switch from disk to SSD.

The bottleneck of write bandwidth is formalized in
a simple equation for predicting
the ideal checkpoint time for a full-memory dump (sustained write)
of all of the
aggregate RAM.  The relationship is well known, and we formalize
it here as the Checkpoint-Fill-Time Law.  We assume here a write
bandwidth (transfer rate) of 100 MB/s for a {\em single} disk.

\def\Storage{{\rm Storage}}
\def\Bandwidth{{\rm Bandwidth}}
\def\Number{{\rm Number}}
\def\SingleDiskFillTime{{\rm SingleDiskFillTime}}
\begin{table*}[ht]
\centering
\scriptsize
\begin{tabular}{|c|c|c|c|c|c|c|c|c|}
  \hline
  Name & Year & $\Storage_{RAM}$ & $\Storage_{d/S}$  & Ratio & Assumed & Assumed single & Single disk & Ideal ckpt \\
       & intro. &            & (disk or SSD)            &    & disk size & disk bandwidth & fill time & time \\
       &  &  &  &  & & & (min.) & (min.) \\
  \hline
  \hline
  Stampede (TACC) & 2014 & 205~TB & 10~PB & 0.02 & 2~TB ?? & 100~MB/s & 333 & 6.7 \\
  \hline
  Jaguar (ORNL) & 2009 & 598~TB & 10.7~PB & 0.056 & 1~TB & 100~MB/s & 167 & 9.4 \\
  \hline
  Titan (ORNL) & 2012 & 710~TB & 10.7~PB & 0.066 & 1~TB & 100~MB/s & 167 & 11.0 \\
  \hline
  Sunway TaihuLight & 2016 & 1,311~TB & ?? & 0.05 ?? & 3~TB ?? & 100~MB/s & 500 & 25.0 ?? \\
  \hline
  CCR (UB) & 2015 & 1.728~TB & 500~TB & 0.0035 & 4~TB & 100~MB/s & 666 & 2.3 \\
  \hline
  SSD-based 4-core node & 2014 & 16~GB & 128~GB & 0.125 & 4~TB ?? & 500~MB/s & 4.3 & 4.3 \\
  \hline
  Theoretical Exascale & 2020 & ?? & ?? & 0.1 ?? & 4~TB ?? & 4~GB/s ?? & 16 & 1.6 ?? \\
  \hline
\end{tabular}
\caption{\label{tbl:ckptFillTime}
	Predictions using the Checkpoint-Fill-Time Law for a full-memory
	dump (checkpoint size = $\Storage_{RAM}$).
	Since the size of disks used by the storage nodes is often not
	reported, it is estimated at half of the largest size disk at
	the time of introduction of the computer.}
\end{table*}

\bgroup
\setlength\tabcolsep{1.5pt} 
\begin{tabular}{rcl}
\medskip
  CkptTime &=& $\displaystyle\Storage_{RAM} \,/\, \Bandwidth_{disks}$ \\
\medskip
             &=& $\displaystyle\Storage_{RAM} \,/\,
	      (\Number_{disks} \times {\rm 100~MB/s})$ \\
\medskip
             &=& $\displaystyle\frac{\Storage_{RAM}}{\Storage_{disks}} \times
	      {\Storage_{disks}\over \Number_{disks}} \,/\, {\rm 100~MB/s}$ \\
\medskip
             &=& $\displaystyle{\Storage_{RAM}\over \Storage_{disks}} \times
              \SingleDiskFillTime_{disk}$ \\
\end{tabular}
\egroup

Similarly, we can write down such a law for an SSD.

\bgroup
\setlength\tabcolsep{1.5pt} 
\begin{tabular}{rll}
\medskip
  CkptTime &=& $\displaystyle{\Storage_{RAM}\over \Storage_{SSDs}} \times
		\SingleDiskFillTime_{SSD}$ \\
\end{tabular}
\egroup

There are many inaccuracies in this law.  As a minor example, this
formula ignores the existence of redundant disks in a RAID configuration.
The aggressive 100~MB/s transfer rate is meant to account for that.
More seriously, a back-end storage subsystem such as Lustre includes
a back-end network that usually cannot support the full bandwidth
of the aggregate disks.  A back-end storage subsystem is optimized
for typical write loads, which are only a fraction of the maximum
write load with all compute nodes writing simultaneously.
Finally, this law is not intended to be used for small checkpoint
images, since this results in small write blocks that are inefficient
for use with disks, and since there is a large variation in
perceived checkpoint time for small writes due to interference
by larger jobs simultaneously using I/O.

Some examples of predictions are shown in Table~\ref{tbl:ckptFillTime}.
The goal of this table is to make a crude prediction on expected
checkpoint times for a future exascale supercomputer based on SSDs.
As will be seen in Section~\ref{sec:evalFullMemory}, the predictions
of the law for Stampede running both HPCG and NAMD
are approximately ten times
faster than what is seen experimentally (after accounting for the
fact that the HPCG computation uses only 1/3 of the available nodes
and only 2/3 of the available RAM per node).

In extrapolating a future exascale SSD-based supercomputer,
the formula predicts a full-memory dump time of 1.6~minutes.
We assume a ratio of RAM to SSD size of 0.1 instead of the
historical 0.02 or 0.05 for disks, since SSD is more expensive
than disk.  It is assumed that SSD will be used for secondary
storage and disk for tertiary storage.
If we accept a factor of ten difference between ideal,
theoretical time and real-world time (in keeping with the approximately
ten-fold penalty seen for HPCG and NAMD in Section~\ref{sec:evalFullMemory},
then this extrapolation predicts a real-world 16~minutes checkpoint
time for exascale computing.

The specification of a future SSD at 4~TB  with
write bandwidth of 4~GB/s is based on an extrapolation from current
SSDs.  At this time, the high-end PCI Express (PCIe) 3.0-based SSDs can achieve
1.5-3.0~TB/s of sequential writes. The PCIe 3.0 interface limits
one today to 8~GB/s.
(The upcoming
PCIe 4.0 promises 16~Gigatransfers/s that typically translate to
16~GB/s.)
We conservatively
assume a bandwidth of 4~GB/s and a 4~TB storage for
for commodity SSDs, four years from now.
This is in keeping with Flash density trends in~\cite[slide~4]{yoon20143d}
and with~\cite{desnoyers2016private}.

\section{Experimental Evaluation}
\label{sec:experiment}

We evaluate our approach for: (a) ability to checkpoint real-world
applications (see section~\ref{sec:evalFullMemory}); scalability
trends across a large range in the number of cores used (see
section~\ref{sec:evalScalability}); and (b) applicability to diverse
environments (see section~\ref{sec:evalDiversity}).

\subsection{Setup}
The experiments were run on the Stampede supercomputer~\cite{stampede}
at TACC (Texas Advanced Computing Center).  As of this writing, Stampede
is the \#10 supercomputer on the Top500 list~\cite{top500}.  In all
cases, each computer node was running 16~cores, based on a dual-CPU
Xeon ES-2680 (Sandy Bridge) configuration with 32~GB of RAM.

Experiments use the Lustre filesystem version 2.5.5 (see
Section~\ref{sec:lustre}) on Stampede.  InfiniBand connections run over
a Mellanox FDR InfiniBand interconnect.  A lower bandwidth Ethernet
connection is available for TCP/IP-based sockets. For all the experiments,
uncompressed memory dumps were used.

The largest batch queue normally available at Stampede provides normally
for 16K~CPU cores, but special permission was obtained to briefly test
at the scale of 32,752~CPU cores. Hence, the maximum scale was 2,047
nodes with 16~CPU cores.

At 32,752~CPU cores, the tests use one-third of the compute nodes of
Stampede.  This is sixteen times as many cores as the largest previous
transparent checkpoint of which we are aware~\cite{cao2014transparent}.
The Stampede supercomputer is rated at 5.2~PFlops (RMAX:sustained) or
8.5~PFlops (RPEAK:peak).~\cite{top500}.  Hence, we estimate our usage
during this checkpoint experiment (using Xeons only) to be a large
fraction of a petaflop.

The experiments at the largest scale were done using a reservation in
which our experiments had exclusive access to up to one-third of the
nodes of Stampede.  The system administrators were careful to monitor our
usage during this period, to ensure that there was no interference with
the jobs of other users.  The administrators observed a peak bandwidth of
116~GB/s to the Lustre filesystem, when we were writing checkpointing image
files at large scale (16K through 32K CPU cores).  The system administrators
of Stampede also observed that they did not find any disturbance in the
workflow of other users during this time.

\subsection{Experiments with Real-world Software}
\label{sec:evalFullMemory}

Two software experiments reflect real-world experience with many
CPU cores and with large
memory footprints.  The High Performance Conjugate Gradients (HPCG)
benchmark represents a realistic mix of sparse and dense linear
algebra~\cite{hpcg-benchmark}, and
is intended to provide a good ``correlation to real scientific
application performance''~\cite{dongarra2013toward}.
The tests with the molecular dynamics simulation NAMD then represents
performance for an application
not based on linear algebra.

\subsubsection{Evaluation of HPCG}

Table~\ref{tbl:hpcg} shows checkpoint and restart times for HPCG at
the scale of 8K, 16K, 24,000, and 32K CPU cores. The 24,000 and 32K core cases
were run with special permission of the system administrators (since at
Stampede, the largest standard batch queue supports only 16K cores),
In all cases, the aggregate size of the checkpoint images per node is 19.2~GB,
representing almost two-thirds of the 32~GB RAM available
on each node.  The bandwidth for writing checkpoints progressively
diminishes with larger size computations, except at the largest scale of
32,752 cores.  This last case was run during a maintenance period,
with presumably little writing associated with that ongoing maintenance.

Note that the Checkpoint Fill-Time Law predicts an ideal
checkpoint time of 6.7 minutes (see Table~\ref{tbl:ckptFillTime}).
At 16K cores, the checkpoint represents 9.4\% of the total
$6,400\hbox{\ nodes}\times 32$~GB, and thus, the observed checkpoint time
is ten times larger than the predicted ideal time of 0.628~minutes.
When pushing Lustre beyond its standard configuration,
the checkpoint for 32,752~cores represents 19.2\% of the total RAM of
$6,400\times 32$~GB, and thus, the observed checkpoint time of 10.9~minutes
is eight times larger than the predicted ideal time of 1.29~minutes.

\begin{table}[ht]
\centering
\begin{tabular}{|c|c|c|c|c|}
  \hline
  \hbox{Num.} of~ & Checkpoint & Restart  & Total ckpt & Write (ckpt) \\
  processes       & time (s)   & time (s) & size (TB)  & bandwidth (GB/s) \\
  \hline
  \hline
  8192 & 136.1 & 215.3 & ~9.4 & 69 \\
  \hline
  16368 & 367.4 & 706.6 & 19 & 52\\
  \hline
  24000 ($^1$) \rule{0pt}{2.6ex}
	 & 634.8 & 1183.8 & 29 & 46 \\
  \hline
  32752 ($^2$) \rule{0pt}{2.6ex}
	& 652.8 & 2539.05 & 38 & 60 \\
  \hline
\end{tabular}
\vskip +6pt
{
\small
\noindent
NOTE: Executed with special permission$^1$; and during Stampede maintenance$^2$
(mostly exclusive access to the cluster). \hfill
}
\caption{\label{tbl:hpcg} Checkpoint and restart trends for HPCG;
  checkpoint image size for each process is 1.2~GB, with 16 images
  generated on each computer node.}
\end{table}

Compared to the checkpoint times, the restarts times are nearly twice
as large. We believe this is because when writing the checkpoint images,
Lustre buffers the checkpoint data. On restart, the checkpoint data needs
to be synchronized to the disk, transferred to each node, and then read
into the memory of each node.

\subsubsection{Evaluation of NAMD}
Table \ref{tbl:namd} shows the results of NAMD for 8K and 16K~CPU cores.
The input parameters are taken from a NAMD-based petascale
study of biomolecular simulations~\cite{phillips2014mapping}.
Comparing these results with those for HPCG, we see that the write
bandwidth for Lustre depends primarily on the number of CPU cores
issuing a sustained write, and does not vary significantly even
though the total checkpoint sizes are smaller for NAMD.
The I/O bandwidths for NAMD correspond roughly with HPCG, and
hence the
actual checkpoint times observed also compare to the ideal times
of Table~\ref{tbl:ckptFillTime} in approximately the same ratio.
(For NAMD with 8K~processes, the 260~MB per process appears to
be a little small for ideal Lustre~I/O.)

\begin{table}[h!t]
\centering
\begin{tabular}{|c|c|c|c|c|}
  \hline
  \hbox{Num.} of~ & Checkpoint & Restart  & Total ckpt & Ckpt (write)  \\
  processes       & time (s)   & time (s) & size (TB)  & bandwidth (GB/s) \\
  \hline
  \hline
  8192 & ~41.4  & 111.4 & 2.1 & 51 \\
  \hline
  16368 & 157.9 & 689.8 & 9.8 & 62 \\
  \hline
\end{tabular}
\vskip +6pt
\caption{\label{tbl:namd} Checkpoint and restart trends for NAMD
  with 8K and 16K cores (one MPI process per core);
  checkpoint image size for each process is 260~MB and 615~MB, respectively,
  with one checkpoint image per process.}
\end{table}

\subsection{Scalability}
\label{sec:evalScalability}

\begin{table}[ht]
\centering
\begin{tabular}{|c|c|}
  \hline
  \hbox{Num.} of~ & Launch   \\
        processes & time (s) \\
  \hline
  \hline
        1K      & 0.3 - 7.5  \\
  \hline
        2K      & 0.8 - 10.5 \\
  \hline
        4K      & 3.2 - 86.7 \\
  \hline
        8K      & 29.2 - 87.9  \\
  \hline
        16K     & 99.3 - 120.8  \\
  \hline
        16K (*) & 15.2 - 21.6  \\
  \hline
\end{tabular}
\vskip +6pt
{
\small
(*) Launch time for 16K processes with checkpointing using
tree of coordinators.
}
\caption{\label{tbl:launch} Launch time for different number of processes running
with checkpointing.}
\end{table}

\begin{figure}[t]
\centering
\includegraphics[width=0.7\columnwidth]{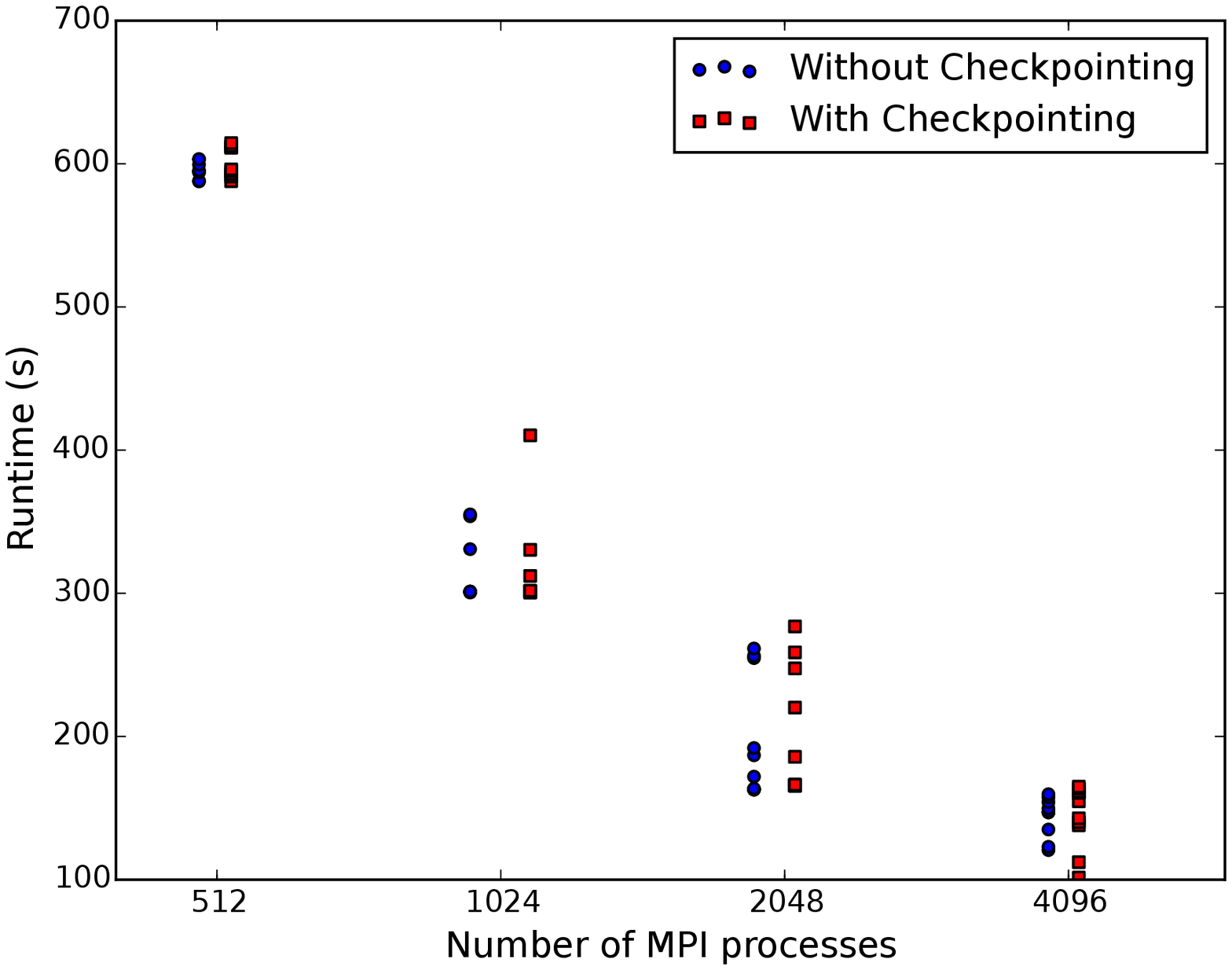}
\caption{\label{fig:runtime-scatter} Runtime overhead for LU.E}
\end{figure}

\subsubsection{Launch Overhead}
\label{subsubsec:launch}

Table~\ref{tbl:launch} shows the overhead incurred when starting the
NAS/NPB LU benchmark~\cite{nas}(NPB~3.3.1, Class E) under our proposed
approach. Recall that the approach uses a centralized coordinator
for coordinating checkpointing among distributed processes and the
coordination messages are sent over TCP/IP network. As a result, we
observe that the launch time scales with the number of processes because
of the increased load on the TCP/IP network. The large variation in the
launch time is due to the congestion on the TCP/IP network that is also
used at Stampede for administrative work. The launch time improves by up
to 85\% at the scale of 16K~processes when we switch to using a tree of
coordinators. Each compute node runs an additional sub-coordinator process that
aggregates and relays requests from the processes on the same node to a
root coordinator. This reduces the network connections by a factor of 16.

\begin{table}[ht]
  \centering
  \begin{tabular}{|c|c|c|c|}
    \hline
    \hbox{Num.} of~ & Runtime (s) & Runtime (s)~~  & Overhead (\%) \\
    processes       & (natively)~~~ & (w/ checkpointing support)   &      \\
    \hline
    \hline
    512 & 596.6 & 601.4 & 0.8  \\
    \hline
    1024 & 316.2 & 317.8 & 0.5  \\
    \hline
    2048 & 197.6 & 201.9 & 2.2   \\
    \hline
    4096 & 144.0 & 144.1 & 0.1  \\
    \hline
  \end{tabular}
  \vskip +6pt
  \caption{\label{tbl:runtime} Runtime overhead for NAS benchmark LU.E (class~E):
Times are native (without checkpointing support) and with checkpointing support.}
\end{table}

\begin{figure*}[t]
  \begin{subfigure}[b]{0.5\textwidth}
    \includegraphics[width=1.0\columnwidth]{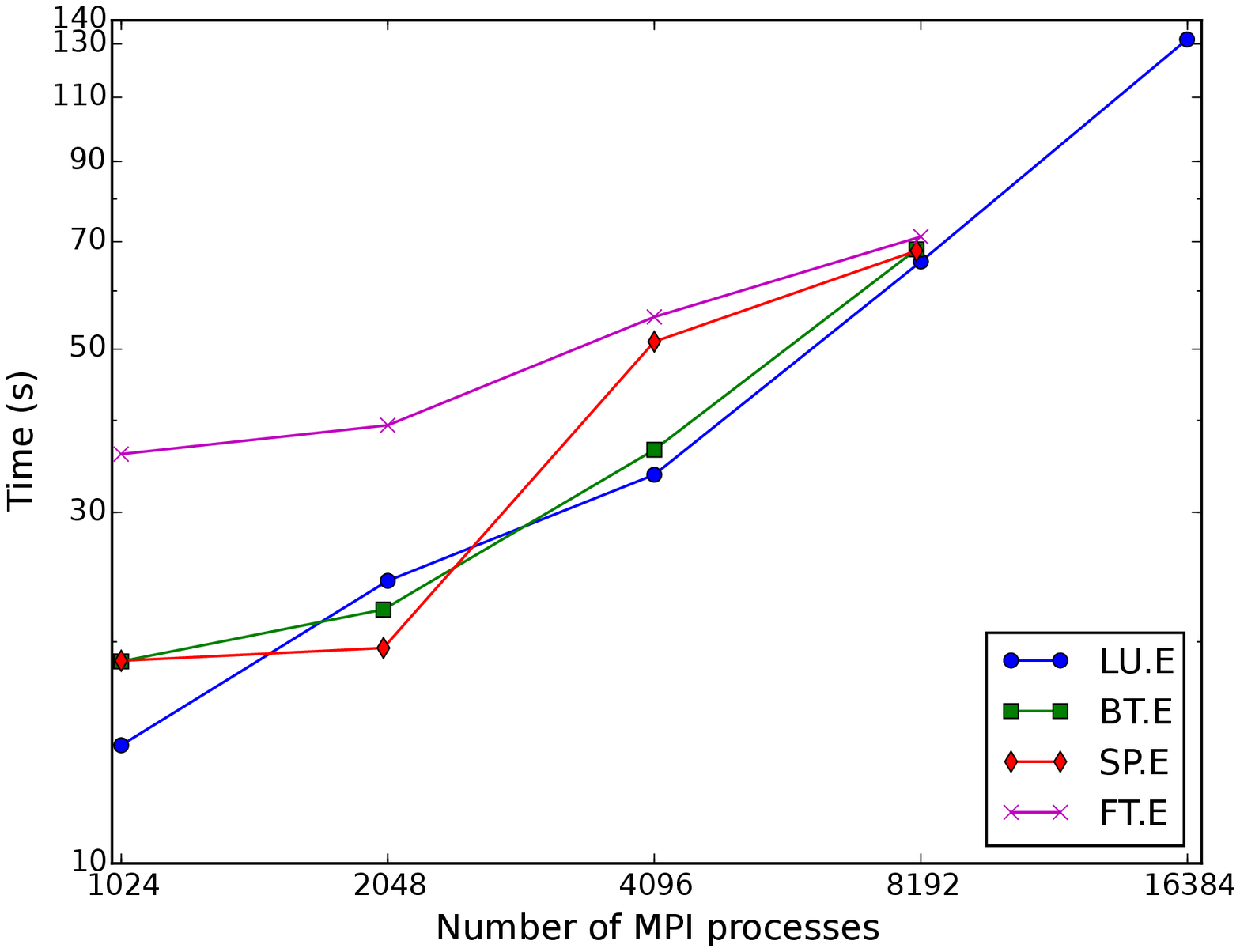}
    \caption{\label{fig:ckpt} Checkpoint Times}
  \end{subfigure}
  \begin{subfigure}[b]{0.5\textwidth}
    \includegraphics[width=1.0\columnwidth]{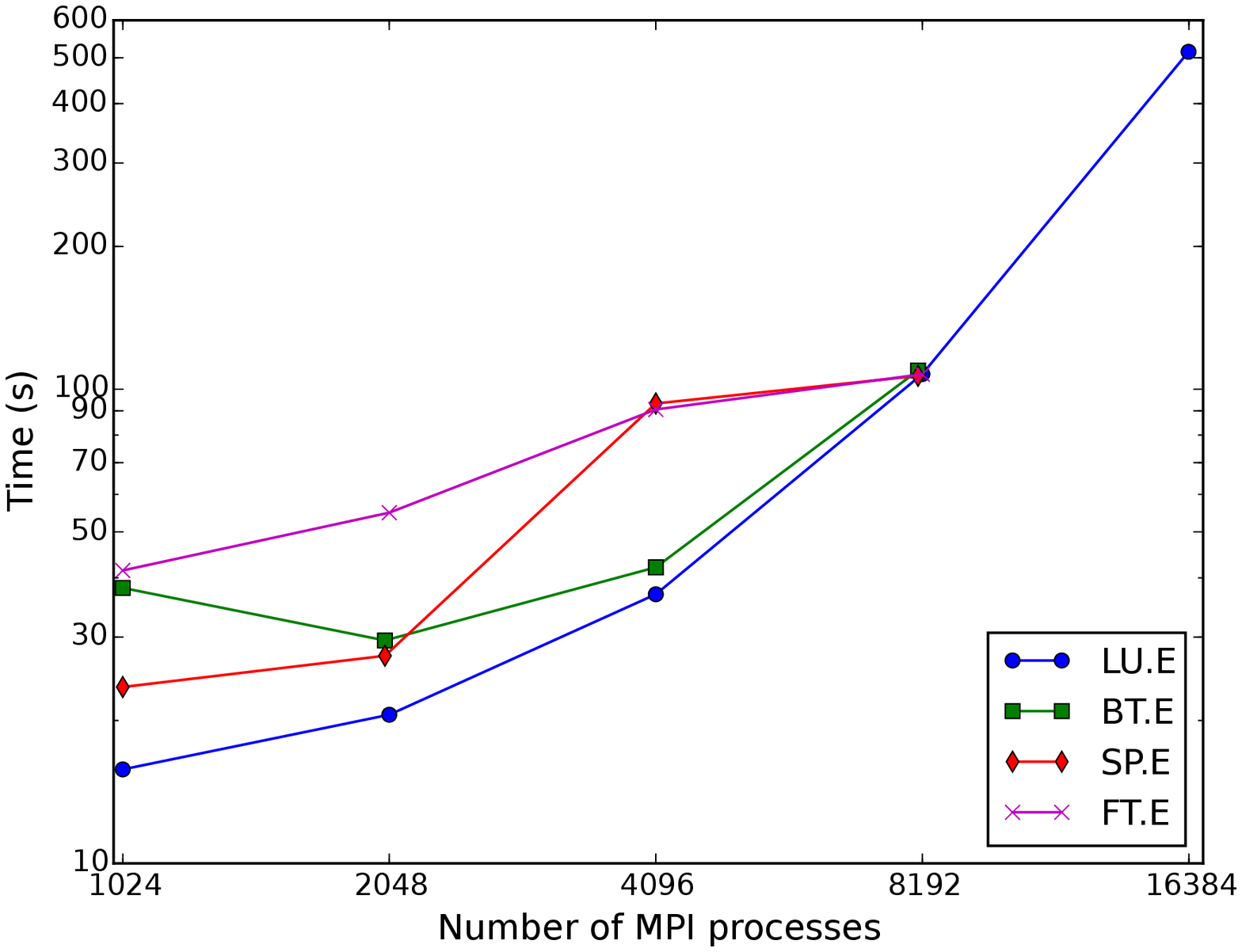}
    \caption{\label{fig:restart} Restart Times}
  \end{subfigure}
  \caption{\label{fig:nas-ckpt-rst} Checkpoint and Restart Times for Four NAS Benchmarks (note the log-log axes)}
\end{figure*}

\subsubsection{Runtime overhead}
\label{subsec:runtime}

Figure~\ref{fig:runtime-scatter} demonstrates the small
overhead of executing with our approach. To minimize the variation in
communication overhead due to network topology, for a given problem
size, the experiments with and without checkpoint support were run on
the same set of nodes.  Even for a fixed set of nodes, we observed a
large variation in the runtimes in successive runs.  We attribute this
to the network congestion on the InfiniBand backend used by Lustre.

Table~\ref{tbl:runtime} shows that the average runtime overhead is less
than 1\% in all cases, except for 2K~processes, where it is 2.2\%. Since
runtime overhead returned to 0.1\% with 4K~processes, we speculate
that the run with 2K~processes suffered from interference by other
users on that day.

As noted in the introduction,
before introducing the optimizations of Section~\ref{sec:runtime-overhead},
we had observed a runtime overhead of 9\% with DMTCP for 4K cores.
The runtime overheads reported here are an important advance.

\begin{table}[ht]
\centering
\begin{tabular}{|c|c|c|c|}
  \hline
  \hbox{Num.} of~ & Checkpoint & Restart  & Checkpoint size     \\
  processes       & time (s)   & time (s) & per process (MB)     \\
  \hline
  \hline
  1024 & 14.5 & 15.8 & 428 \\
  \hline
  2048 & 24.2 & 20.6 & 342 \\
  \hline
  4096 & 33.7 & 36.9 & 300 \\
  \hline
  8192 & 65.8 & 107.6 & 280 \\
  \hline
  16368 & 131.8 & 514.7 & 285 \\
  \hline
\end{tabular}
\vskip +6pt
\caption{\label{tbl:lue} Checkpoint and restart trends for LU.E}
\end{table}

\subsubsection{Checkpoint-Restart Trends}

The scalability trends up to 16K~CPU cores are demonstrated for the NAS
LU benchmark in Table~\ref{tbl:lue}.

The checkpoint overhead can be divided into two parts: the communication
between the compute processes and the central coordinator, and the
work to write the checkpoint images. Since there are only a few small
messages that are sent to coordinate checkpointing, the checkpoint time
is dominated by the time Lustre takes to write the checkpoint images.

Notice that the time to write a checkpoint
image differs by up to 99\% at the scale of 16K processes, even though
the sizes of their images are the same. This difference increases with
the number of processes. We attribute this to Lustre creating and writing
meta-data for each file.

The restart overhead also consists of two parts: the time to build the
connection with the coordinator, and the time to read into the memory
the checkpoint images.  In the case of restart, it is the work to build
all the connections that dominates the trend. It follows the same trend
as in Section~\ref{subsubsec:launch}, since it uses the same model to
build the connections.

\subsection{Diversity}
\label{sec:evalDiversity}

In this section, we demonstrate the support for different applications
as well as different MPI implementations. Apart from LU, we test three
other NAS benchmarks: BT, SP, and FT. In addition, we test the Scalable
Molecular Dynamics (NAMD) real-world application~\cite{namd}.
We also show the support for two
other popular MPI implementations: Open~MPI~\cite{openmpi} and
Intel~MPI~\cite{intelmpi}.

\subsubsection{Evaluation of NAS/NPB Benchmarks}

The performance of BT, SP, and FT is demonstrated for up to 8k~CPU cores.
Together with LU, the checkpoint times and the restart times are shown
in Figures~\ref{fig:ckpt} and~\ref{fig:restart}, respectively.

The memory overhead of the proposed approach is negligible compared
with the memory footprint of the application. As a result, for a given
scale the checkpoint times and the restart times roughly correspond the
checkpoint image sizes, regardless of the application type.

\subsubsection{Evaluation of Different MPI Implementations}
\label{subsec:ompiAndImpi}

\begin{table}[h!t]
\centering
\begin{tabular}{|c|c|c|c|}
  \hline
  \hbox{MPI}      & Checkpoint & Restart  & Checkpoint size     \\
  implementation  & time (s)   & time (s) & per process (MB)     \\
  \hline
  \hline
  Intel MPI & 298.9 & 191.8 & 775 \\
  \hline
  Open MPI  & 299.7  & 128.5  & 520 \\
  \hline
\end{tabular}
\vskip +6pt
\caption{\label{tbl:mellanox} Comparison of two MPI implementations;
         LU.E for 500 processes}
\end{table}

The experiments with Open~MPI and Intel~MPI were done at the Center
for Computational Research at the State University of New York,
Buffalo~\cite{buffalo}. The results shown in Table~\ref{tbl:mellanox}
demonstrate that the proposed approach is MPI-agnostic. While the
restart times roughly correspond to checkpoint image sizes, the
checkpoint times don't. This is attributed to filesystem backend caching.

\begin{table*}[ht]
\scriptsize
  \begin{subtable}[ht]{0.32\textwidth}
\centering
\begin{tabular}{|c|c|c|c|}
  \hline
  \hbox{Num.} of~ & Ckpt & Rst  & Ckpt size     \\
  processes       & time (s)   & time (s) & (MB)     \\
  \hline
  \hline
  1024 & 18.8 & 38.1 & 675 \\
  \hline
  2025 & 22.1 & 29.5 & 480 \\
  \hline
  4096 & 36.5 & 42.1 & 368 \\
  \hline
  8100 & 68.3 & 109.5 & 331 \\
  \hline
\end{tabular}
\vskip +6pt
\caption{\label{tbl:bte}BT.E}
\end{subtable}
\begin{subtable}[ht]{0.32\textwidth}
\centering
\begin{tabular}{|c|c|c|c|}
  \hline
  \hbox{Num.} of~ & Ckpt & Rst  & Ckpt size     \\
  processes       & time (s)   & time (s) & (MB)     \\
  \hline
  \hline
  1024 & 18.9 & 23.5 & 634 \\
  \hline
  2025 & 19.6 & 27.4 & 452 \\
  \hline
  4096 & 51.28 & 93.3 & 368 \\
  \hline
  8100 & 68.0 & 106.7 & 332 \\
  \hline
\end{tabular}
\vskip +6pt
\caption{\label{tbl:spe}SP.E}
\end{subtable}
\begin{subtable}[ht]{0.32\textwidth}
\centering
\begin{tabular}{|c|c|c|c|}
  \hline
  \hbox{Num.} of~ & Ckpt & Rst  & Ckpt size     \\
  processes       & time (s)   & time (s) & (MB)     \\
  \hline
  \hline
  1024 & 36.0 & 41.5 & 1200 \\
  \hline
  2048 & 39.4 & 54.9 & 703 \\
  \hline
  4096 & 55.3 & 90.6 & 488 \\
  \hline
  8192 & 71.1 & 107.5 & 385 \\
  \hline
\end{tabular}
\vskip +6pt
\caption{\label{tbl:fte}FT.E}
\end{subtable}
\caption{\label{tbl:spec}Checkpoint and restart trends for various NAS
benchmarks}
\vskip -6pt
\end{table*}

\section{Discussion of Scalability Issues}
\label{sec:discussion}

\subsection{Centralized Coordinator}

Our approach uses a single-threaded central coordinator.  It appears that this
design does {\em not} insert a central point of contention.  In these
experiments, the checkpoint coordinator was always run on a separate compute
node with no competing processes.  Total network traffic on each socket
was estimated to be a total of 20~KB during a checkpoint-restart.  This
traffic was primarily related to the publish-subscribe database
maintained by the InfiniBand checkpointing code.  Nevertheless, the CPU
load was always measured at less than 5\% of the time for one CPU core.

Separately, the approach uses TCP sockets to communicate
with the peer processes.  This represents a design flaw at the
petascale level.  Two issues were encountered.
First, the use of multiple TCP writes without an intervening read
forced us to
invoke {\tt TCP\_NODELAY} to turn off Nagle's algorithm~\cite{nagle}.
Second, there was a need at larger scale to use a staggered sleep
(network backoff) during
initialization of TCP sockets, so that the many peers would
not overwhelm the operating system (or possibly the switch hardware)
in a burst of requests to create new socket connections.

Additionally, most Linux-based operating systems include a limit
on the number of socket connections per process.  Our implementation needed to
be extended with a tree-of-coordinators so that the many peers
connecting to the coordinator would not exceed this limit.
It is in part for this reason that MVAPICH2 switches to an on-demand
connect mode when more than 64~processes are present.

\subsection{Ethernet and IPoIB}

A related issue was encountered specifically for Stampede.  While a single
10~Gb~Ethernet backbone is provided for each compute rack, that network
is often overloaded.  Because it is required for the administration
of the Stampede supercomputer, administrative measures are sometimes
taken to kill processes that excessively use the Ethernet network for
communication.

In an effort to alleviate this,
we experimented with IP over InfiniBand (IPoIB)~\cite{ipoib} for all
TCP communication, in order to avoid
using the Ethernet network.  The IPoIB implementation at Stampede is based
entirely on a kernel module, and does not expose the underlying
InfiniBand layer in the user space.  For some unknown reason, we found
that the Stampede system continued to kill processes when there
were too many socket connections (albeit now of the IPoIB flavor).
Also, it was observed that IPoIB
at Stampede used the second, lower-bandwidth port of the HCA InfiniBand
adapter.  Hence, IPoIB provided better latency, but did not provide
better bandwidth than TCP over Ethernet.

\subsection{Filesystem Backend}

Initially, we were worried whether the Lustre filesystem
could handle the large bandwidth during checkpoint.
As related in Section~\ref{sec:experiment}, the use of 24,000 CPU cores
was allowed on Stampede only through a special reservation.
System administrators monitoring the backend performance reported that
there was no measurable interference with other concurrent users.

\subsection{Better support from the InfiniBand device drivers}

As discussed in~\ref{sec:runtime-overhead}, the shadow send/receive queues
provide stronger correctness guarantees but impose a significant runtime
overhead.  The proposed alternative is to use a heuristic-based approach
with relaxed correctness guarantees. A third alternative is possible if
the InfiniBand device driver can provide an API to ``peek'' into the
hardware to learn the current state of the send/receive queues. While
being non-destructive, the peek operation could significantly simplify
the logic around draining and refilling of the send/receive queues without
imposing a runtime overhead.

\subsection{Fast Restart using Demand-paging}
During restart, there is an opportunity to use ``mmap'' to map the
checkpoint image file back into process memory (RAM) on-demand.  Instead,
all memory was copied from the checkpoint image file to process memory
(RAM) during restart.  With ``mmap'', the restart could be significantly
faster for a certain class of application that have a smaller working set.
This would allow for some overlap of computation and demand-paging
generated file~I/O.  Further, there is less of a ``burst'' demand on the
Lustre filesystem.  This mode was not used, so that the worst-case time
for restart could be directly measured.

\section{Related Work}
\label{sec:related}

To the best of our knowledge, the largest previous checkpoint was carried out
by Cao~\hbox{et al.}~\cite{cao2014transparent}.  That work demonstrated
transparent checkpoint-restart over InfiniBand~RC (but not UD~mode)
for the first time.
Scalable results were demonstrated for the NAS NPB LU benchmark for
2048~MPI processes over 2048~CPU cores.  That work mostly used local
disk rather than Lustre, showing a maximum I/O bandwidth of 0.05~GB/s
when using the local disks of 128~nodes.
(One example with Lustre over 512~processes reported an I/O bandwidth
of 0.1~GB/s.)
The previous work was demonstrated solely for Open~MPI using RC~mode,
while today most MPI implementations also take advantage of
InfiniBand UD~mode during initialization.

The most frequently used packages for system-level transparent
checkpointing today are
BLCR~\cite{BLCR03,BLCR06},
CRIU~\cite{criu},
Cryopid2~\cite{cryopid2}, and
DMTCP~\cite{ansel2009dmtcp}.
Only DMTCP and BLCR are used for checkpointing MPI computations.
DMTCP is the only one of the four that supports transparent
checkpointing of distributed computations, and so it supports
our current MPI-agnostic approach toward checkpointing.

In contrast, BLCR is also often used for checkpointing MPI, but
only in combination with
an MPI-specific checkpointing service such
as \cite{hursey2009interconnect} for Open~MPI or
\cite{sankaran2005lam} for LAM/MPI.
BLCR can only checkpoint the processes on a single node.
Hence, an MPI-specific checkpointing service
temporarily tears down the InfiniBand network, and then uses
BLCR~\cite{BLCR03,BLCR06} to checkpoint individual nodes as
standalone computations.  Afterwards, the InfiniBand connections
are re-built.

DMTCP is preferred over the combined use of BLCR with an MPI
implementation-specific checkpointing service for two reasons: (a)~It is
MPI-agnostic, operating without modification for most MPI implementations;
and (b)~the alternative checkpointing service that tears down the network
can incur long delays when re-initializing the InfiniBand connections
upon resuming the computation and hence limits its performance.

In 2014, Cao \hbox{et al.}~\cite{cao2014transparent} extended the DMTCP
model from transparent support for TCP to include InfiniBand using the
then dominant RC communication mode.  That work was demonstrated for
Open~MPI~1.6 --- mostly checkpointing to the local disk.  They found
that checkpointing to the Lustre filesystem was 6.5~times faster than without
Lustre, although restart times were similar in the case of LU.E over
512~CPU cores.  Most of that work was done with the DMTCP default
of ``on-the-fly'' gzip compression of checkpoint images, and with
checkpointing to local disk with 2048~CPU cores.

There have been several surveys of the state of the art for
software resilience in the push to petascale and then exascale
computing~\cite{elnozahy2004checkpointing,
	snir2014addressing,
	cappello2009fault,cappello2009toward,cappello2014toward}.
One of the approaches is FTC-Charm++~\cite{zheng2004ftc}, which
provides a fault-tolerant runtime base on
an in-memory checkpointing scheme (with a disk-based extension)
for both Charm++ and AMPI
(Adaptive MPI).  Three categories of checkpointing are
supported:  uncoordinated, coordinated, and communication-induced.

Because of the potentially long times to checkpoint, a multi-level
checkpointing approach~\cite{moody2010design} has been proposed.
The key idea is to support local fault tolerance for the ``easy'' cases,
so that a global checkpoint (potentially including a full-memory dump)
is used as a last resort.  Since restart from a global checkpoint are
needed less often, such checkpoints may also be taken less often.

A popular application-level or user-level mechanism
is ULFM (user-level failure mitigation).  By applying recovery
at the user-level, they offer different recovery models,
such as backward versus forward, local versus global and
shrinking versus non-shrinking.
\cite{laguna2016evaluating} reviews the ULFM model, and adds
an application-level model based on global rollback.

Finally, rMPI (redundant MPI) has been proposed for exascale
computing~\cite{ferreira2011rmpi}.   This has the potential
to make checkpointing less frequent, and thus allow for longer
times to checkpoint.  The authors write, ``Note that redundant
computing $\ldots$ reduces the overhead of checkpointing but does not
eliminate it.'' The authors provide the example of a fully-redundant
application for which Daly's equation~\cite{daly2006higher} predicts
a run for 600~hours without failure over 50,000~{\em nodes} with
a 5-year MTTI/node.~\cite[Figure~12]{ferreira2011rmpi} (MTTI is
mean-time-to-interrupt.)

\section{Conclusion}
\label{sec:conclusion}

The need for a fault-tolerance solution for exascale computing has been
a long-time concern~\cite{elnozahy2004checkpointing,cappello2009fault,
cappello2009toward,cappello2014toward,snir2014addressing}.  This work has
demonstrated a practical petascale solution, and provided
evidence that the approach scales into the exascale generation.
Specifically, system-initiated full-memory dumps for three modern MPI
implementations over InfiniBand have been demonstrated.  This required
virtualization of InfiniBand~UD, since the previous simpler
InfiniBand~RC point-to-point mode did not support modern
MPI implementations at scale.

Testing on real-world-style applications of NAMD and HPCG stressed
large memory footprints.  The current Lustre filesystems successfully
supported many-terabyte full-memory dumps.  A simple formula in
Section~\ref{sec:ssd} allowed for extrapolation to future SSD-based
exascale computers.  The predicted ideal checkpoint time was 1.7~minutes,
which extrapolates to under 17~minutes (ten-fold increase) after comparing
the ideal formula against current supercomputers.

In particular, special permission was received to run HPCG with 32,752 CPU
cores (one-third of the Stampede supercomputer), and a 38~TB checkpoint
image was created in 10.9~minutes.  The system administrator manually
monitored a similar run with 24,000 cores, and reported that it did not
affect the normal use of I/O by other concurrent users.

\section*{Acknowledgment}
We would like to acknowledge the comments and encouragement of Raghu Raja
Chandrasekar in integrating DMTCP with MVAPICH2.  We would also like
to acknowledge Zhixuan Cao, Peter Desnoyers,
and Shadi Ibrahim for helpful feedback.
We also acknowledge
the support of the Texas Advanced Computing Center (TACC) and the
Extreme Science and Engineering Discovery Environment (XSEDE), which
is supported by National Science Foundation grant number ACI-1053575.
We especially would like to acknowledge the assistance of Tommy Minyard
and Bill Barth from TACC in helping set up and monitor the large
experiment there.  Also, we acknowledge the resources at the University
at Buffalo Center for Computational Research (www.buffalo.edu/ccr)
that were used in performing a portion of the numerical experiments.

\end{document}